\documentclass[fleqn,12pt,twoside]{article}
\usepackage{espcrc1}

\usepackage{graphicx}
\usepackage[figuresright]{rotating}

\newcommand{\AmS}{{\protect\the\textfont2
  A\kern-.1667em\lower.5ex\hbox{M}\kern-.125emS}}

\title{$\gamma$-Radiation of thermalized Quark-Gluon-Plasma}

\author{Haitham Zaraket \address[MCSD]{Physics department\\ 
        The University of Winnipeg\\
        515 Portage Avenue\\
        Winnipeg, Manitoba, Canada\\
        R3B 2E9}}
\begin{document}

\maketitle

\begin{abstract}
Long time ago, photon production was proposed as 
a probe and a thermometer 
for Quark-Gluon Plasma (QGP). However, 
only recently has the complete $\alpha_s$ order photon spectrum been obtained.
In this paper we give a brief review of the problematic as well 
as discuss the ${\cal O}(\alpha_s)$ result. 
\end{abstract}

\section{Why Photon production?}
Following the well known {\it rhetoric}, photons are weakly coupled to the
strongly  interacting quarks and gluons. Photons emitted in a QGP   will
immediately escape without further interactions in the plasma. This immediate
``escape'' will carry useful information on the  nature of the supposedly
formed plasma, at the emission stage. Hence the comparison  between the {\em
calculated} photon spectrum from a QGP and that from a hadron gas  medium with
the measured photon  spectrum in heavy ion collision, after background
subtraction, will constitute  an evidence in favor of either state of matter.\\
In an optimistic scenario, the plasma will live long enough for thermalization 
to occur. At a suitable high temperature ($T$), which is not so realistic in present 
heavy ion collisions, the running strong coupling constant 
$\alpha_s$ will be small.\\
The above framework can be summarized as follows: we have a system in thermal 
equilibrium with an exact microscopic description in terms of quarks and gluons. In the 
small coupling constant regime, 
the calculation of the photon spectrum seems to be a straightforward 
application of perturbation theory. This is seemingly a simple situation  compared to 
photon production in proton-proton collision where form-factors show up. 
Although photons are weakly coupled to the plasma this description is 
oversimplified, since the quark that emits the photon 
is affected by medium effects such as Debye screening; it also acquires a thermal 
mass which plays a central role in screening infrared divergences. Medium 
effects are well described by the Hard Thermal Loop (HTL) effective theory 
\cite{BraatP1}. So, the HTL theory is  the natural scheme for calculating 
the photon production rate in a thermalized QGP. \\
For present (RHIC), and near future (LHC), heavy ion collision experiments 
a more realistic prediction should go beyond the small coupling constant regime. 
This will be briefly discussed at the end of this article.

\section{A survey}
In a QGP in local thermal equilibrium, the photon production rate or the number of 
photons, with momentum $Q=(q_0,{\bf q})$, produced 
per unit time and per unit volume is conveniently expressed in terms of the 
imaginary part of the photon two point function, calculated in the plasma 
\begin{equation}
{{dN^{\gamma}}\over{dtd{\bf  x}}}=
-{{d{\bf q}}\over{(2\pi)^4 2q_o}} {2n_{_{B}}(q_o)}\; {\rm Im}\,
\Pi{}(q_o,{\bf q})\; .
\end{equation}
This formula is valid at lowest order in the electromagnetic coupling, and to all 
orders in the strong coupling constant. In other words, the two point function encodes 
the essential physics of the plasma, relevant for photon emission.\\
As mentioned in the previous section, to obtain the photon two point function
in a perturbative procedure, the HTL theory should be used. The one loop 
calculation in the effective theory was performed some  time ago
\cite{KapusLS1,BaierNNR1}. It was believed that the one loop  result 
gave {\bf {the}} leading order $\alpha_s$ photon spectrum.   This belief was based
on the success of the effective theory. The essence of the  effective theory is
to reorder the perturbation theory to give a meaning 
to the  equivalence
between the loop-wise expansion and the successive orders in the  coupling
constant. The loophole in this equivalence  
comes from collinear  configurations.
The  effective theory does not incorporate a ready-to-use remedy for collinear 
problems. Collinearity is kinematical,  {\it i.e.} it depends on the particular
problem of interest, so it is not  surprising if the effective theory fails to
treat such problems systematically. \\
\begin{figure}
\centerline{\resizebox*{!}{3cm}{\includegraphics{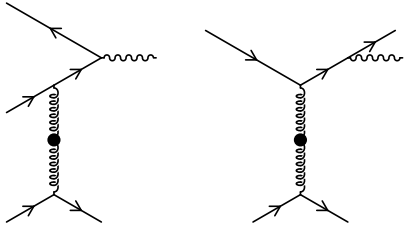}}}
\caption{\label{fig:processes} The bremsstrahlung and off-shell
annihilation processes that are expected to dominate the 
the production of hard photons and low mass hard dileptons.}
\end{figure}
It worth mentioning that the underlying physical processes in the one loop 
diagram are the 
$2\leftrightarrow 2$ processes, Compton and annihilation. However, it was 
shown \cite{AurenGKP1,AurenGKP2,AurenGKZ1} that bremsstrahlung and the new process 
of annihilation of an 
off-shell quark (on the left of figure \ref{fig:processes}) contribute to leading 
order. These processes appear only at the two loop level in the effective 
theory. The breakdown of the effective theory, 
two loop is of the same order as one loop, is traced back 
to the appearance of collinear divergences 
({\em {i.e.}} the collinear emission of the photon by a quark). 
This collinearity is soften by the use 
of the effective theory which gives a kinematical cutoff denoted by 
\begin{equation}
M_{\rm eff}^2=M_\infty^2+\frac{Q^2}{q_0^2}p_0(p_0+q_0)\; ,
\label{eq:Mass}
\end{equation}
where $M_\infty\sim gT$ is the thermal mass of fermions, provided by the effective 
theory, and  $p_0$ is the quark energy. The present form of the cut-off is valid 
for real and 
small mass virtual photons (the real photon case is obtained for $Q^2=0$). 
The dependence 
of this cut-off on the coupling constant, although it regularizes what would be a 
collinear singularity, renders the two loop diagrams equally important 
as the one loop contribution.
Since, after phase space integration the cut-off $M^2_\infty$ appears linearly
in the denominator, its $g^2$ dependence cancels the extra powers of the
couplings in the two-loop diagrams.\\
It is thus natural to ask whether this breakdown of the effective theory does not 
propagate to higher loop orders? \\ 
Simple power counting \cite{AurenGZ2} indicates that the collinear configuration 
persists for all ladder diagrams. As a consequence, higher loop diagrams do contribute 
to the leading $\alpha_s$ order for real photon production. Hence a 
resummation of a whole set of gauge invariant diagrams is mandatory.
  
\section{The physics of different scales}
\begin{figure}
\centerline{\resizebox*{!}{3cm}{\includegraphics{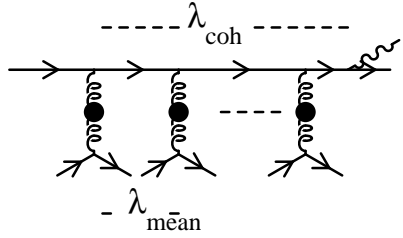}}}
\caption{\label{fig:LPM} Rescattering of a quark where the 
formation time ($\sim \lambda_{\rm coh}$) is larger than the mean free 
path. }
\end{figure}
In attempt to focus on the underlying physical mechanism which renders 
higher loop diagrams equally important as the one loop diagram, we calculated 
\cite{AurenGZ3} 
the imaginary part of the photon polarization tensor at two loop level with quarks 
having a finite width. This study 
leads to a simple physical picture. Photon emission 
from a quark gluon plasma gives rise to two natural physical scales 
(figure \ref{fig:LPM}):
\begin{itemize}
\item the formation length (time) $l_F$, 
also called the coherence length $\lambda_{\rm coh}$:
\begin{equation}
l_F^{-1}\sim \delta E= \frac{q_o}{2p_0(p_0+q_0)}\left[p_\perp^2+M_{\rm
eff}^2\right]\;, 
\end{equation}
where $p_\perp$ is the quark transverse (compared to photon momentum) momentum, 
\item the mean free path (width)$^{-1}$: $\lambda_{\rm mean}\sim (g^2T\ln(1/g))^{-1}$
\end{itemize}
The interplay between these two scales leads to the emergence of different 
physical regimes:
\begin{itemize}
\item the perturbative regime ($l_F < \lambda_{\rm mean}$), where the dominant mechanisms
of photon production are Compton and annihilation appearing at one loop  order 
in the HTL effective theory. High dilepton mass spectrum is a typical example 
of this regime. 
\item the Landau-Pomeranchuk-Migdal regime ($l_F\gg\lambda_{\rm mean}$): it is the 
region where the formation length is much longer than the mean free path. Inelastic 
processes like bremsstrahlung should be considered. This leads to coherent photon 
production, which is responsible for the LPM suppression of the photon spectrum. 
In this region a further resummation is needed to include rescattering. Real photon 
production exemplifies this regime.
\end{itemize}   

\section{Resummation} 
The finite width study cited in the previous section,
although it provides a nice  physical picture, is far from giving the complete
photon spectrum. A correct approach would consist in
re-summing the ladder diagrams mentioned above together with quark
propagators with a finite width (self-energy corrections). This has been
recently carried out \cite{ArnolMY1,ArnolMY2}. The authors of
\cite{ArnolMY1,ArnolMY2} showed that important cancellations of 
long ranged interactions occured between
vertex and self-energy diagrams and derived an integral 
equation with a simple physical interpretation. The bottom line of their resummation is 
to consider the rescattering 
of an almost on-shell quark in a random Gaussian background field. The imaginary 
part of the two point function was found to be:
\begin{eqnarray}
&&\!\! {\rm Im}\,\Pi_{_{R}}{}_\mu^\mu(Q)\approx {{e^2
N_c}\over{2\pi}}
\int_{-\infty}^{+\infty}dp_0\,
\;[n_{_{F}}(p_0+q_0)-n_{_{F}}(p_0)]\;
{{p_0^2+(p_0+q_0)^2}\over{2(p_0(p_0+q_0))^2}}
\nonumber\\
&&\qquad\qquad\qquad\times{\rm Re}\,\int {{d^2{\bf p}_\perp}\over{(2\pi)^2}}\;
{\bf 
p}_\perp\cdot{\bf f}({\bf p}_\perp) \nonumber
\label{eq:AMY}
\end{eqnarray}
with 
\begin{equation}
$$2{\bf p}_\perp=i\delta E {\bf f}({\bf p}_\perp)+g^2 C_{_{F}}T\int
{{d^2{\bf l}_\perp}\over{(2\pi)^2}} {\cal C}({\bf l}_\perp)
[{\bf f}({\bf p}_\perp)-{\bf f}({\bf p}_\perp+{\bf l}_\perp)]\; ,
\end{equation}
$\delta E$ is the same as the inverse of the formation time defined before. This 
integral equation reflects the following features: 
\begin{itemize}
\item The combination 
$[{\bf f}({\bf p}_\perp)-{\bf f}({\bf p}_\perp+{\bf l}_\perp)]$ 
guarantees the 
cancellation of all infrared behavior ({\it a priori} non-perturbative) when 
$l_\perp\leq g^2T $. 
\item Iteration is equivalent to rescattering in the medium with a collision 
term ${\cal C}$. 
\item The width, discussed in the previous section, could be seen as a part of the 
collision term. We notice again that multiple scatterings are important when the 
collision integral is of the same order as the $\delta E$ term, as was predicted 
by the model discussed in the previous section. 
\end{itemize}
The above equation is solved numerically and it is found that rescattering leads 
to about 25 \% suppression of the photon spectrum coming from bremsstrahlung and 
off-shell scattering. Hence, although the ``next to one loop'' 
processes are not as enhanced as was thought originally \cite{AurenGKZ1} they still 
give important contribution which should be included beside the strict one loop 
processes to obtain  the complete order 
$\alpha_s$ photon spectrum. \\
Recently the collision term was obtained analytically using new sum rules at 
finite temperature \cite{AurenGZ4}. The collision term is found to be 
\begin{equation}
{\cal C}({\bf l}_\perp)={1\over{l_\perp^2}}-{1\over{l_\perp^2+3m_{\rm g}^2}}\; .
\end{equation}
This analytical form allows to circumvent the evaluation of complicated integrals  
in the original form of the collision term. It is also useful for the extension of the above 
study to the production of low mass dileptons. 

\section{Conclusions}
Photon production nowadays is known to complete order $\alpha_s$. It is among 
the rare theoretical calculations going beyond the leading logarithm approximation. 
The resummation done by Arnold {\it et al} leads to a picture where one averages over
a Gaussian background field which goes beyond the classics of static scattering 
centers extensively used in the literature to model rescattering.\\
It must be emphasized that bremsstrahlung and off-shell annihilation are among 
the dominant mechanisms for photon production. Although the resummation has led to 
a 25 \% suppression, the next to one loop processes still give 
significant contributions which will enhance the photon production rate in a
quark-gluon plasma.

\section{Extensions} 
Thermal photon production is not the unique source of photons.
Pions, for  example, decay into photons giving a very important background. This
background  should be  subtracted in order to isolate the thermal photon
production and compare it to 
photon production in a hadron gas model. On the other hand,  dilepton
production has a different background to be subtracted. Hence a compilation  of
photon and dilepton will constitute a tractable mean for ``plasma
detection''.\\ 
The resummation done in \cite{ArnolMY1,ArnolMY2} includes only
the transverse photon  polarization. However, dilepton could  receive
contributions from the  longitudinal polarization sector, hence the extension
of the above  resummation to  dilepton case requires some precautions. The
necessary resummation for low mass  dilepton leads to  \cite{AurenGMZ1}
(preliminary results) 
\begin{equation}
{\rm Im}\,\Pi_{_{R}}{}_\mu^\mu(Q)\approx {e^2}\int dp_0 [\dots]\int 
\left[2{\bf p}_{\perp}.{\bf f}({\bf p}_{\perp})\oplus Q^2 g({\bf p}_{\perp})\right]
\end{equation}
the new scalar function $g$ satisfies an equation analogue to that satisfied 
by ${\bf f}$. The preliminary results for the dilepton rate indicates an LPM type  
suppression  which does not rule out the contribution coming from bremsstrahlung 
and off-shell annihilation found recently in \cite{AurenGZ5}.\\
We should stress that the small coupling constant regime is an idealized situation. 
A more realistic 
coupling constant should not be so small. Recall that photon emission is dominated 
by collinear emission, this occurs for almost on-shell quarks. Hence quarks  
can be treated to a good approximation in a quasi-particle model, with masses 
derived from lattice calculation.  This is under investigation \cite{Gelis}.

\end{document}